\documentclass{ws-procs9x6}

\begin{document}

\title{MESONS AND NUCLEONS FROM HOLOGRAPHIC QCD}

\author{ULUGBEK YAKHSHIEV$^*$ and HYUN-CHUL KIM}

\address{Department of
Physics, Inha University, Incheon 402-751, Korea\\
$^*$E-mail: yakhshiev@inha.ac.kr}

\author{YOUNGMAN KIM}

\address{Asia Pasific Center
for Theoretical Physics and Department of Physics\\ Pohang
University of Science and Technology, Pohang, Gyeongbuk 790-784,
Korea}

\begin{abstract}
We present in this talk a recent investigation on a unified approach
within the framework of a hard-wall model of AdS/QCD. We first
study a theoretical inconsistency in existing models. In order to
remove this inconsistency, we propose a unified approach in which the
mesons and the nucleons are treated on the same footing, the same
infrared (IR) cutoff being employed in both fermionic and bosonic
sectors. We also suggest a possible way of improving the model by
introducing a five-dimensional anomalous dimension.
\end{abstract}


\bodymatter

\section{Introduction}\label{Intr}
The AdS/QCD model has been extensively used in describing a wide variety
of phenomena in particle and nuclear physics and even in condensed
matter physics. The
model~\cite{Erlich:2005qh,DaRold:2005zs,DaRold:2005vr} is
constructed as a 5D holographic dual of QCD based on the general
wisdom of
AdS/CFT~\cite{Maldacena:1997re,Gubser:1998bc,Witten:1998qj}, the 5D
gauge coupling being identified by matching the two-point vector
correlation functions in deep Euclidean region.  Although the
AdS/QCD model is based on an \textit{ad hoc} ground, it reflects
some of most important features of gauge/gravity duality. Moreover, it
is rather successful in describing properties of mesons~(see, for
example, the following recent review~\cite{Erdmenger:2007cm}).

QCD is not a conformal theory, in particular, in
the low-energy region, so one should also incorporate this property
in constructing an effective AdS/QCD model.  Consequently, different
models have been developed. In
Refs.~\cite{Erlich:2005qh,DaRold:2005zs,DaRold:2005vr}, the size of
the extra dimension (also known as the compactification scale) was
fixed at the point that corresponds approximately to the QCD scale
parameter $\Lambda_{\rm QCD}$, i.e., an infrared (IR) cutoff
parameter was explicitly introduced.  It is often considered as
the confinement scale that also breaks sharply the conformal
invariance. These AdS/QCD models are called the \textit{hard-wall
model}.  On the contrary, there is an alternative approach called a
\textit{soft-wall model} in which the conformal invariance is broken
smoothly by introducing the dilaton-like field in the 5D AdS
space~\cite{Batell:2008zm,Batell:2008me,Andreev:2006vy}.

While both approaches describe meson properties relatively well, one
confronts a serious problem in the baryonic sector.  Since there is no
theoretical reason that the confinement scale for the baryon should be
the same as that for the meson, different values of the IR cutoff for
the baryon have been
introduced~\cite{Contino:2004vy,Gherghetta:2000qt,Mueck:1998iz,
Hong:2006ta,Arutyunov:1998ve,Henningson:1998cd}. Actually, the main
reason for that is due to the fact that it is impossible to reproduce
the meson and baryon properties, in particular, excited nucleon
states~\cite{Hong:2006ta}, with the same value of the IR
cutoff used. However, when one calculates the meson-nucleon coupling
constants, a serious inconsistency arises~\cite{Maru:2009ux}.  In
order to determine the coupling constants consistently, one must use
the \textit{same} IR cutoff.  Otherwise, one cannot fully consider
whole information on the meson and nucleon wavefunctions.  This is the
motivation of the present investigation and has been studied in our
recent work~\cite{HCKim09}.  In the present talk, we will briefly
review how to resolve this inconsistency and will put forward possible
methods to describe the meson and baryon on an equal footing.

\section{The effective action of a hard-wall model with holographic
  mesons and nucleons}
We start from a hard-wall model for mesons developed in
Refs.\cite{Erlich:2005qh,DaRold:2005zs} and study its applications
to nucleons\cite{Hong:2006ta,HCKim09}. The model has a geometry of
5D AdS
\begin{equation}
ds^2 = g_{MN} dx^M dx^N \;=\; \frac1{z^2}(\eta_{\mu\nu}dx^\mu
dx^\nu-dz^2)\,,
\end{equation}
where $\eta_{\mu\nu}$ stands for the 4D Minkowski metric:
$\eta_{\mu\nu} = \mathrm{diag}(1,-1,-1,-1)$. The 5D AdS space is
compactified by two different boundary conditions, i.e. the IR
boundary at $z=z_m$ and the UV one at $z=\epsilon\to 0$. Considering
the global chiral symmetry $\mathrm{SU(2)}_L\times\mathrm{SU(2)}_R$
of QCD, we need to introduce 5D local gauge fields $A_L$ and $A_R$
of which the values at $z=0$ play a role of external sources for
$\mathrm{SU(2)}_L$ and $\mathrm{SU(2)}_R$ currents respectively.
Since chiral symmetry is known to be broken to $\mathrm{SU(2)}_V$
spontaneously as well as explicitly, we introduce a bi-fundamental
field $X$ with respect to the local gauge symmetry
$\mathrm{SU(2)}_L\times\mathrm{SU(2)}_R$, in order to realize the
spontaneous and explicit breakings of chiral symmetry in the AdS
side. Considering these two, we can construct the bi-fundamental 5D
bulk scalar field $X$ in terms of the current quark mass $m_q$ and
the quark condensate $\sigma$
\begin{equation}
X_0(z)=\langle X \rangle \;=\; \frac12(m_q z + \sigma z^3)
\end{equation}
with isospin symmetry assumed.

The 5D gauge action in AdS space with the scalar bulk field and the
vector field  can be expressed as
\begin{eqnarray}
S_M &=&\int d^4x\int dz\,  {\cal L}_M\,,\nonumber\\
{\cal L}_M&=&\frac1{z^5}\, \mathrm{Tr} \left[ |DX|^2 +3|X|^2
-\frac{1}{2g_5^2}(F_L^2 + F_R^2) \right], \label{eq:eff_meson}
\end{eqnarray}
where covariant derivative and field strength tensors are defined as
$DX =\partial X - i A_L X + iXA_R$ and $F_{L,R}^{MN}=\partial^M
A_{L,R}-\partial^N A_{L,R}-i[A^M_{L,R},\,A^N_{L,R}] $.  The 5D gauge
coupling $g_5$  is fixed by matching the 5D vector correlation
function to that from the operator product expansion (OPE):
$g_5^2=12\pi^2/N_c$.  The 5D mass of the bulk gauge field $A_{L,R}$
is determined by the relation $m_5^2=(\Delta - p)(\Delta + p
-4)$\cite{Gubser:1998bc,Witten:1998qj} where $\Delta$ denotes the
canonical dimension of the corresponding operator with spin $p$ and
turns out to be $m_5^2=0$ because of gauge symmetry. The effective
action~(\ref{eq:eff_meson}) describes the mesonic
sector\cite{Erlich:2005qh,DaRold:2005zs} completely apart from
exotic mesons\cite{Kim:2008qh}.

To consider baryons in the flavor-two ($N_F=2$) sector, one needs to
introduce a bulk Dirac field corresponding to the nucleon at the
boundary\cite{Hong:2006ta,HCKim09}. This hard-wall model was applied
to describe the neutron electric dipole moment\cite{Hong:2007tf} and
holographic nuclear matter\cite{Kim:2007xi}.  In this model, the
nucleons are the massless chiral isospin doublets $(p_L,\, n_L)$ and
$(p_R,\,n_R)$ which satisfy the 't Hooft anomaly matching. Then the
spontaneous breakdown of chiral symmetry induces a chirally
symmetric mass term for nucleons
\begin{equation}
\mathcal{L}_{\chi SB} \sim -M_N \left( \begin{array}{c} \bar{p}_L \\
    \bar{n}_L  \end{array}\right)\, \Sigma\, (p_R, \,n_R) + \mathrm{h.c.},
\end{equation}
where $\Sigma=\exp(2i\pi^a\tau^a/f_\pi)$ is the nonlinear
pseudo-Goldstone boson field that transforms as $\Sigma\to U_L
\Sigma U_R^\dagger$ under $\mathrm{SU(2)}_L\times\mathrm{SU(2)}_R$.
The $\tau^a$ and $f_\pi$ represent the SU(2) Pauli matrices and the
pion decay constant, respectively. Thus, we have to consider the
following mass term in the AdS side
\begin{equation}
\mathcal{L}_{\mathrm{I}} = -g \left( \begin{array}{c} \bar{p}_L \\
    \bar{n}_L  \end{array}\right)\, X\, (p_R, \,n_R) + \mathrm{h.c.},
\end{equation}
where $g$ denotes the mass coupling (or Yukawa coupling) between $X$
and nucleon fields, which is usually fitted by reproducing the
nucleon mass $M_N=940$ MeV. In this regard, we can introduce two 5D
Dirac spinors $N_1$ and $N_2$ of which the Kaluza-Klein (KK) modes
should include the excitations of the massless chiral nucleons
$(p_L,\, n_L)$ and $(p_R,\,n_R)$, respectively.  By this
requirement, one can fix the IR boundary conditions for $N_1$ and
$N_2$ at $z=z_m$.

Note that the 5D spinors $N_{1,2}$ do not have chirality. However,
one can resolve this problem in such a way that the 4D chirality is
encoded in the sign of the 5D Dirac mass term. For a positive 5D
mass, only the right-handed component of the 5D spnior remains near
the UV boundary $z\to 0$, which plays the role of a source for the
left-handed chiral operator in 4 dimension. It is vice versa for a
negative 5D mass.  The 5D mass for the $(d+1)$ bulk dimensional
spinor is determined by the AdS/CFT expression
\begin{equation}
(m_5)^2 \;=\; \left(\Delta -\frac{d}{2}\right)^2 \label{eq:5dmass}
\end{equation}
and turns out to be $m_5=5/2$. However, since QCD does not have
conformal symmetry in the low-energy regime, the 5D mass might
acquire an anomalous dimension due to a 5D renormalization flow.
Though it is not known how to derive it, we will introduce some
anomalous dimension to see its effects on the spectrum of the
nucleon.

Summarizing all these facts, we are led to the 5D gauge action for
the nucleons
\begin{eqnarray}
S_N &=& \int d^4x\int dz\, \frac1{z^5}\, \mathrm{Tr} \left[{\cal
    L}_{\mathrm{K}}+{\cal L}_{\mathrm{I}} \right], \cr
{\cal L}_{\mathrm{K}} &=&  i \bar{N}_1
 \Gamma^M \nabla_M N_1 + i \bar{N}_2
 \Gamma^M \nabla_M N_2
 - \frac52 \bar{N}_1 N_2 +  \frac52 \bar{N}_2 N_1
\nonumber\\
{\cal L}_{\mathrm{I}} &=& -g\left[ \bar{N}_1 X N_2 + \bar{N}_2
  X^\dag N_1\right],
\label{EffAct}
\end{eqnarray}
where
\begin{equation}
 \nabla_M = \partial_M + \frac{i}{4}\, \omega_M^{AB} \Gamma_{AB}
 -i A_M^L\,.
\end{equation}
The non-vanishing components of the spin connection are
$w_M^{5A}=-w_M^{A5}=\delta_M^A/z$ and
$\Gamma_{AB}=\frac1{2i}[\Gamma^A,\Gamma^B]$ are the Lorentz
generators for spinors.  The $\Gamma$ matrices are related to the
ordinary $\gamma$ matrices as $\Gamma^M= (\gamma^\mu,\,-i\gamma_5)$.

The more details of the present approach can be found in
Ref.\cite{HCKim09}.

\section{Results and discussions}

Before presenting the results of this work, we note the most of
input parameters of the model such as $m_q$, $\sigma$ and $z_m$ are
quite well fitted in the mesonic sector~\cite{Erlich:2005qh}. Hence,
we have only one free parameter $g$ to reproduce the data in the
baryonic sector. However, the IR cutoff $z_m$ in the baryonic
sector, which is often interpreted as a scale of the confinement,
takes different values from those in the mesonic sector.  Actually,
Ref.~\cite{Hong:2006ta} performed two different fittings of these
parameters. In the first fitting of Ref.~\cite{Hong:2006ta}, the
$z_m$ and the $\sigma$ were fixed in the mesonic sector, and the $g$
is fitted to the nucleon mass. In the second fitting, the $z_m$ and
the $g$ were taken respectively to be $(205\,\mathrm{MeV})^{-1}$ and
$14.4$ such that the masses of the nucleon and the Roper resonance
$N(1440)$ were reproduced. Since there is no reason for a nucleon to
have the same scale of the confinement as that for a meson, this
might be an acceptable argument as far as one treats mesons and
baryons separately. However, there is one caveat. When it comes to
some observables such as the meson-baryon coupling constants, we
need to treat the mesons and baryons on the same footing and require
inevitably a common $z_m$. Otherwise, we are not able to consider
whole information on both mesons and baryons. Moreover, a model
uncertainty brings on by the mass coupling $g$. Thus, in the present
section, we will carry out the numerical analysis very carefully,
keeping in mind all these facts.

We first take different values of the $z_m$ from those in the
mesonic sector and try to fit the data as was done in
Ref.~\cite{Hong:2006ta}.  In this case, $\sigma$ is defined as
$\sigma={4\sqrt2}({g_5z_m^3})^{-1}$. In general, one can examine two
different limits of the mass coupling $g$ (see Ref.~\cite{HCKim09}):
In the limit of the small mass coupling, there are three free
parameters $m_q$, $g$ and $z_m$. All other parameters can be related
to $z_m$. On the other hand, in the limit of the strong mass
coupling, the $g$ can be fixed by some condition~\cite{HCKim09},
which leaves only two free parameters. Obviously, the dependence on
the current quark mass $m_q$ must be tiny because of its smallness,
so we can simply neglect it.  In this case, we have only one free
parameter.

\begin{table}
\tbl{The results of the spectra of the nucleon and the $\rho$ meson.
In the limit of the small mass coupling, there are three free
parameters $m_q$, $g$ and $z_m$, while in the limit of the strong
mass coupling, the $g$ is fixed near its critical
value~\cite{HCKim09}.  All dimensional quantities are expressed in
units of MeV. The asterisks indicate input parameters. The parameter
$\sigma$ is defined as $\sigma={4\sqrt2}({g_5z_m^3})^{-1}$. The 5D
nucleon mass $m_5$ is given by the AdS/CFT, Eq.~(\ref{eq:5dmass}). }
{\begin{tabular}{@{}cccccccccc@{}}\toprule $m_5$& $z_m^{-1}$ &
$\sigma^{1/3}$ &$m_q$& $g$&
$(p,n)^+$ & $N^+$ &$N^-$&$\rho$&$\rho$\\
&&&&&(939)&(1440) &(1535)&(776)&(1475)\\
\colrule \multicolumn{10}{c}{The limit of the  strong mass coupling}\\
5/2& $130.9^*$& 126.4 & $0$ & 15.9& 940 &1336  &1367 &315&723\\
5/2&$129.7^*$ & 125.2 & $3^*$  & 15.7& 940 & 1328 & 1358&312&716\\
5/2&$126.0^*$ & 121.7 & $10^*$  & 15.4& 940 & 1305 & 1332&303&696\\
\colrule \multicolumn{10}{c}{The case of the small  mass coupling}\\
5/2& $147.2^*$& 142.1 & $0$ & 6.0$^*$& 940 &1440  &1457 &354&813\\
5/2&$147.0^*$ & 141.9 & $3^*$  & 6.0$^*$& 940 & 1439 & 1456&354&812\\
5/2&$146.3^*$ & 141.3 & $10^*$  & 6.0$^*$& 940 & 1435 & 1451&352&808\\
\botrule
\end{tabular}}
\label{tabH}
\end{table}

The results of the calculations are listed in table~\ref{tabH}.  In
the upper part of the table, we present the results in the limit of
the strong mass coupling. They are more or less the same as those
obtained in Ref.~\cite{Hong:2006ta}.  For comparison, we list the
results for the small mass coupling in the lower part of
table~\ref{tabH} where the mass coupling $g$ is chosen to be $6$.
While the spectrum of the nucleon seems to be qualitatively well
reproduced, that of the $\rho$ meson is fairly underestimated in
comparison with the experimental data. In the case of the strong
mass coupling, the situation becomes even worse. One can note
however, as shown in Ref.~\cite{HCKim09}, the ordering of the
nucleon-parity states are correctly reproduced for $0<g<g_{\rm
crit}$.

The results listed in table~\ref{tabH} indicate that it is not
possible to reproduce the spectra of the $\rho$ meson and the
nucleon at the same time.\footnote{Note that in the present work we
do not aim at the fine-tuning of the parameters to reproduce the
experimental data. The output data in baryonic sector is quite
stable for changes in $\sigma$.} As an attempt to improve the
above-presented results, we want to introduce an anomalous dimension
of the 5D nucleon mass while the 5D mass of the bulk vector field
does not acquire any anomalous dimension because of the gauge
symmetry.

\begin{figure}
\begin{center}
\psfig{file=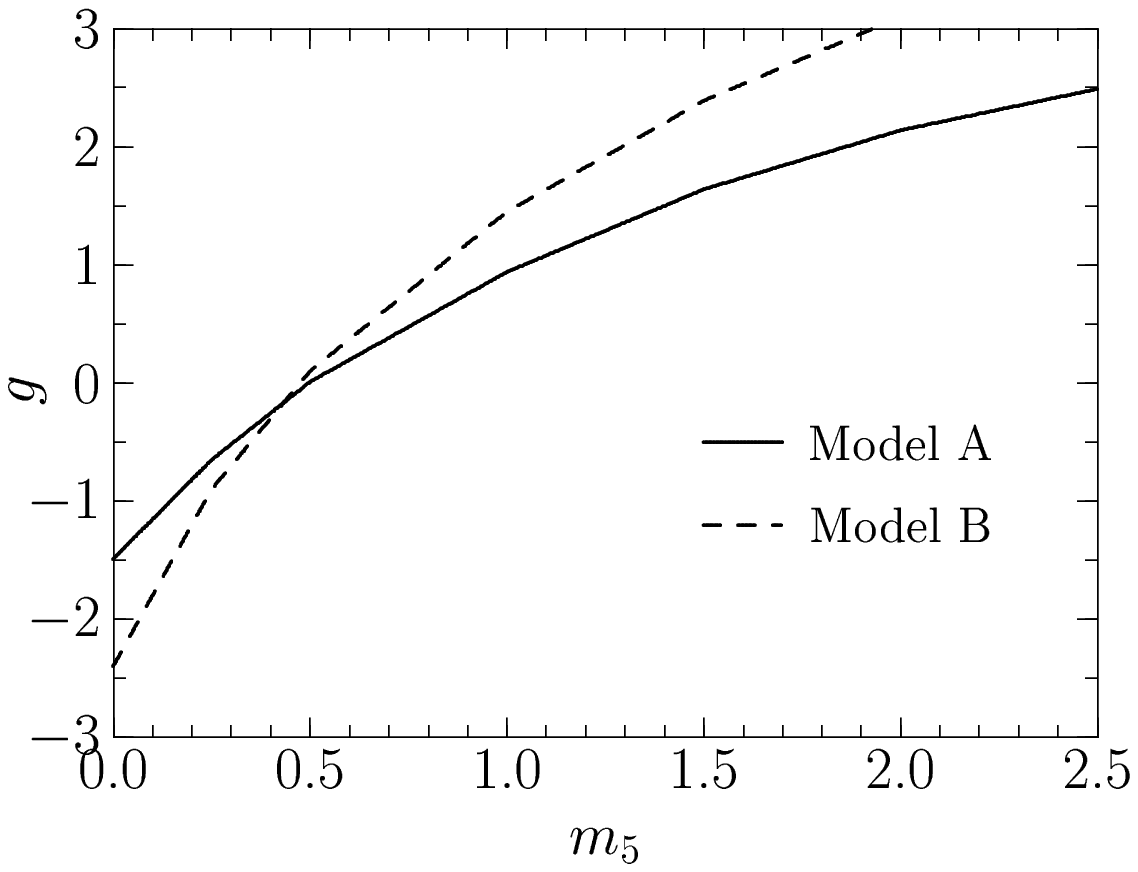,width=3.5 in}
\end{center}
\caption{Mass coupling $g$ dependence on the renormalized 5D mass
$m_5$. The parameters for models A (solid curve) and B (dashed one)
are taken from mesonic sector (see
Ref.\protect\cite{Erlich:2005qh}).} \label{fig1}
\end{figure}

\begin{figure}
\begin{center}
\psfig{file=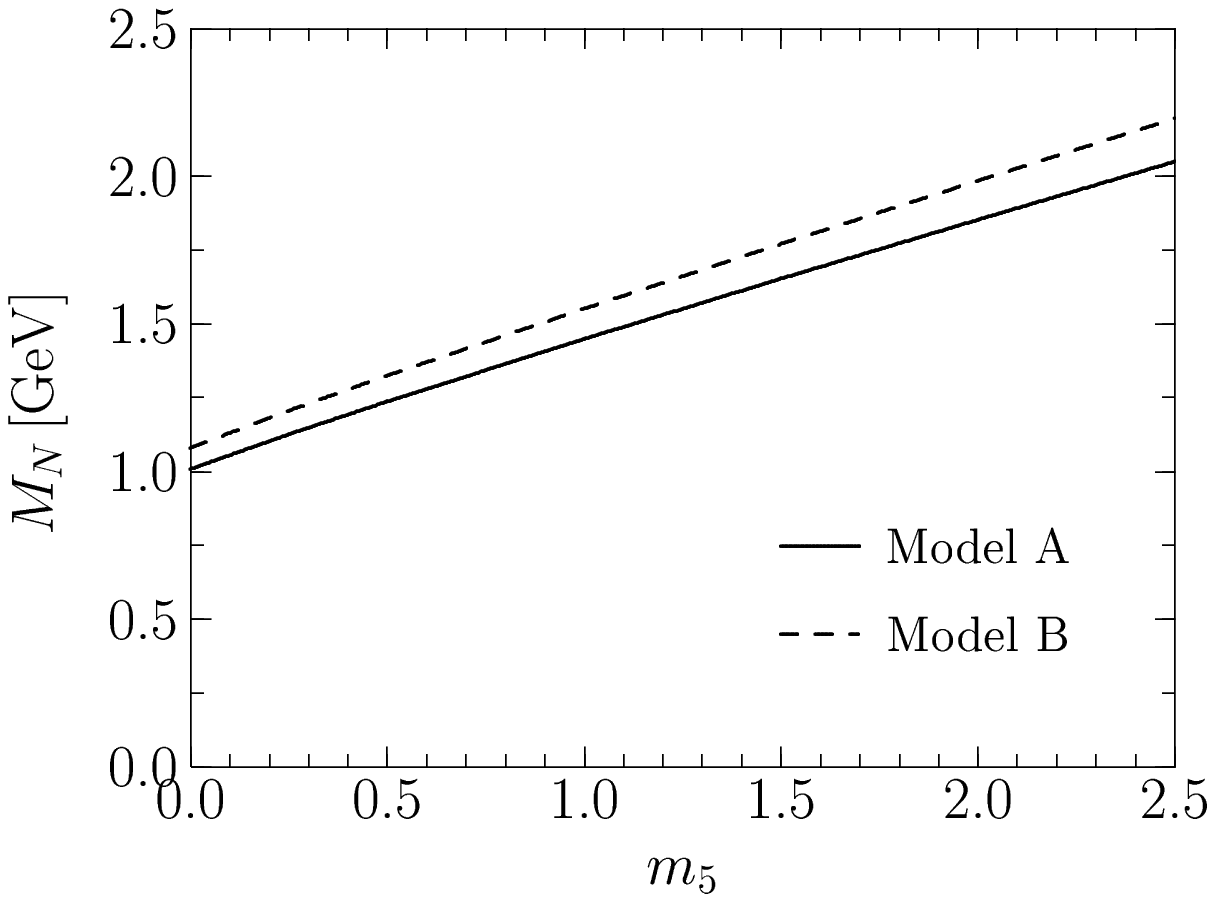,width=3.5 in}
\end{center}
\caption{The mass of the lowest-lying nucleon (in GeV) as a function
of the renormalized 5D mass $m_5$. The notations are same as in
fig.\ref{fig1}.} \label{fig2}
\end{figure}

The results are drawn in figures~\ref{fig1} and~\ref{fig2}. One can
see that, when anomalous dimension is set to zero (i.e. $m_5=5/2$ is
fixed), the experimental data is badly reproduced. The inclusion of
an additional parameter (i.e. considering $m_5$ as a free parameter)
improves the output data well but leads to larger values of the
possible anomalous dimension.

Note that here the nucleon mass is not used as an input.  Varying
the value of $m_5$, we try to fit the spectrum of the nucleon. We
present the results from two different parameter sets called model A
and model B. In this analysis, we take the values of the $z_m$ and
$\sigma$ from Ref.~\cite{Erlich:2005qh}.  Note that the $\rho$ meson
mass is used as an input in model A, while model B corresponds to
the global fitting done in Ref.~\cite{Erlich:2005qh}. The 5D nucleon
mass is varied in the range of $0\le m_5\le 5/2$, its anomalous
dimension being considered as mentioned before. As shown in
figures~\ref{fig1} and~\ref{fig2}, the best result is obtained with
$m_5=0$. Though the absolute value of the nucleon mass turn out to
be overestimated in contrast to the previous analysis presented in
table~\ref{tabH}, qualitatively it is well reproduced within
$30\,\%$ while $\rho$ meson mass is fitted around its experimental
value.

We are now in a position to include meson-baryon coupling constants
in the present numerical analysis. We will consider here the $\pi
NN$ and the $\rho NN$ coupling constants in addition to the $\rho$
meson and the nucleon spectra.  One has to keep in mind that in
order to calculate the meson-baryon coupling constants it is
essential to use the same $z_m$ for the mesonic and baryonic
sectors. Otherwise, it is not possible to keep whole information on
the wavefunctions. Thus, it is of utmost importance to compute all
observables with the same set of parameters. We perform a global
fitting procedure to obtain the results listed in
table~\ref{globfitG}.  Note that we consider here the chiral limit
($m_q=0$), since its effects on the results are rather
tiny.\footnote{Note that in the chiral limit, the nucleon mass is
different from experiments, $M_n\simeq 939~{\rm MeV}$. For instance,
$M_n\simeq 882~{\rm MeV}$ in the chiral limit~\cite{PMWHW}.}
\begin{table}
\tbl{The results of the spectra of the nucleon and the $\rho$ meson
and the $\pi NN$ and $\rho NN$ coupling constants.  The parameters
$z_m$, $\sigma$, and $g$ are found by the global fitting procedure.
The anomalous dimension of the 5D nucleon mass is chosen in such a
way that the 5D mass vanishes. All other definitions are the same as
in table~\ref{tabH}.} {\begin{tabular}{@{}cccccccccc@{}}\toprule
$z_m^{-1}$ &$\sigma^{1/3}$  &$g$& $(p,n)^+$ & $N^+$
&$N^-$&$\rho$&$\rho$&$g_{\pi NN}$
&$g_{\rho NN}$\\
&&&(939)&(1440)&(1535)&(776)&(1475)&(13.1)&(2.4)\\
\colrule
$285^*$ & $256^*$&   -2.0$^*$ &890 & 1791 & 1797&685&1573&1.65&1.39\\
$285^*$ & $237^*$&    -2.0$^*$ &890 & 1790 & 1796&685&1573&1.76&1.39\\
$285^*$ & $256^*$&   -8.0$^*$ &930 & 1826 & 1856&685&1573&4.89&1.34\\
$285^*$ & $237^*$&   -8.0$^*$ &920 & 1817 & 1843&685&1573&5.12&1.35\\
\colrule
$285^*$ & $227^*$&   -9.6$^*$ &930 & 1826 & 1856&685&1573&6.12&1.34\\
\colrule
$280^*$ & $252^*$&  -2.0$^*$ &874 & 1760 & 1765&673&1546&1.65&1.39\\
$280^*$ & $233^*$&  -2.0$^*$ &874 & 1759 & 1764&673&1546&1.76&1.39\\
 \botrule
\end{tabular}}
\label{globfitG}
\end{table}
We assume also that the 5D nucleon mass acquires a large anomalous
dimension so that it may vanish, i.e., $m_5=0$.  The best fit is
obtained with the parameters fitted as follows:
$z_m=(285\,\mathrm{MeV})^{-1}$, $\sigma=(227\,\mathrm{MeV})^3$, and
$g=-9.6$.  The masses of the ground-state nucleon and the $\rho$
meson are in good agreement with the experimental data. Moreover,
those of the excited states are qualitatively well reproduced within
$10-20\,\%$. However, the coupling constants are in general about
$50\,\%$ underestimated.  We mention that in Ref.~\cite{Maru:2009ux}
the dependence of the meson-baryon coupling constants on the $z_m$
was investigated without considering hadron spectra but the results
for the coupling constants are more or less in the same level as in
the present work.
\section{Summary and outlook}
We have investigated the mesons and the nucleons in a unified
approach, based on a hard-wall model of
AdS/QCD~\cite{Erlich:2005qh,Hong:2006ta}. In order to study the
nucleon spectrum, we have developed an approximated method in which
the effective potential can be expanded. The method of this
approximation was shown to work very well.  In particular, the correct
ordering of the nucleon parity states was \textit{analytically} as well
as \textit{explicitly} shown in this method.~\cite{HCKim09}

We then have performed several numerical analyzes, varying the model
parameters such as the IR cutoff $z_m$, the quark condensate
$\sigma$, and the mass coupling (or Yukawa coupling) $g$. In order
to improve the results of the nucleon and the $\rho$ meson spectra
on an equal footing, we have introduced an anomalous dimension of
the 5D nucleon mass. We found that the zero 5D nucleon mass,
$\Delta=2$, produces the best fit.

We have proceeded to compute the $\pi NN$ and $\rho NN$ coupling
constants with the same IR cutoff $z_m$ used. This is crucial in order
to keep whole information about the wavefunctions.  We carried out
the global fitting procedure in which we obtained the best fit with
the values of the parameters:
$z_m=(285\,\mathrm{MeV})^{-1}$, $\sigma=(227\,\mathrm{MeV})^3$, and
$g=-9.6$. The mass spectra of the nucleon and the $\rho$ meson are
in relatively good agreement with the experimental data within
$10-20\,\%$, whereas the $\pi NN$ and $\rho NN$ coupling
constants underestimated by about $50\,\%$.

Last but not least, we want to mention the following significant
point. While the present AdS/QCD model for the baryon respects some
important physics such as spontaneous chiral symmetry breaking, it
still suffers from a serious flaw. Any construction of AdS/QCD models
must satisfy the UV matching with QCD. However, the present framework
of the AdS/QCD model for the baryon does not match the OPE results
of QCD. It indicates that there is still a very important component
missing in the present version of the model, that is, the correct
surface terms in the 5D effective action are missing. The
corresponding investigation is under way~\cite{Kimetal}.

\section*{Acknowledgments}

U.T. Yakhshiev would like to thank the organizers of HNP09 for the
possibility to give a talk and for the hospitality during his stay
at Osaka University. The present work is supported by Basic Science
Research Program through the National Research Foundation of Korea
(NRF) funded by the Ministry of Education, Science and Technology
(grant number: 2009-0073101). Y. Kim acknowledges the Max Planck
Society(MPG), the Korea Ministry of Education, Science,
Technology(MEST), Gyeongsangbuk-Do and Pohang City for the support of
the Independent Junior Research Group at the Asia Pacific Center for
Theoretical Physics(APCTP).

\end{document}